\documentclass[]{elsart}
\usepackage[]{epsfig}
\begin{document}

\begin{frontmatter}

\title{Cornell potential in Kalb-Ramond scalar QED
via Higgs mechanism}

\author{Anais Smailagic 
}
\address{ INFN, Sezione di Trieste,\\
          Trieste, Italy}
\author{Euro Spallucci  \thanksref{s}}
\thanks[s]{e-mail : euro@ts.infn.it}
\address{Dipartimento di Fisica, Gruppo Teorico, Universit\`a di Trieste,
and INFN, Sezione di Trieste,\\
Trieste, Italy}

\begin{abstract}
In this letter we derive the Cornell confining potential in a theory of interacting
Abelian gauge vector and massive Kalb-Ramond tensor. The Kalb-Ramond mass
is instrumental to obtain the linear confining behavior of the potential at large
distances. The same model can be described via interaction with Higgs fields, alter-
natively, providing mass to the vector, or to the tensor fields. In the first case, the
photon acquires mass, while the tensor remains massless. The resulting interaction
potential is a screened Coulomb one. In
the second case, the photon remains massless while Kalb-Ramond tensor acquires
mass and the resulting potential is of the Cornell type with the mass parameter
determining the slope of the linear part.
\end{abstract}

\end{frontmatter}

\section{Introduction}
 
QCD is the widely accepted theoretical framework successfully describing
Strong Interactions at large momenta. On the other hand, the experimental
evidence indicates that the low momenta ( large distance ) regime is char-
acterized by the confinement of the hadronic elementary constituents , i.e.
quarks and gluons. So far, it has not been possible to obtain in a clear and
unambiguous way a linear confining potential. Nevertheless, there is a general
consensus about the Abelian character of the confinement phenomenon 
\cite{Luscher:1978rn,Kondo:1997pc,Kondo:1997kn,Kondo:2014sta}. 
 In the meantime the confinement is currently simulated by the phenomeno-
logical class of potentials commonly known as Cornell-type potentials 
\cite{Eichten:1978tg}, consisting in its simplest version of a sum of an attractive Coulomb potential and
a linearly increasing part.
In this letter we will present an Abelian toy-model giving rise to the above
mentioned Cornell potential. In order to achieve this goal one has to suitably
modify standard QED which by itself gives only the Coulomb interaction be-
tween static charges.
As a guiding line one considers Lee-Wick type modification of Maxwell elec-
trodynamics originally motivated by the desire to eliminate short-distance
singularity inherent in the Coulomb potential 
\cite{Lee:1969fy,Lee:1970iw,Accioly:2010js,Accioly:2011zz}. 
In fact, it turns out that
the short distance behavior is linear while at large distance the Coulomb tail
is recovered.
Lee-Wick model actually leads to a desired “ confining ” behavior but in the
wrong range of distances. Therefore, we shall introduce a modification of the
Lee-Wick Lagrangian in a way to flip short and long distance behavior.
To indicate qualitatively the way to switch the two regimes, let us start from
the Lee-Wick modification of QED by adding a higher derivative term to the
Maxwell Lagrangian

\begin{equation}
 \mathcal{L}_{LW}=-\frac{1}{4}F_{\mu\nu}F^{\mu\nu}+\frac{1}{4}F_{\mu\nu}\frac{\partial^2}{m^2}F^{\mu\nu} \label{lw}
\end{equation}

The second term in (\ref{lw}) dominates at short distance and leads the linear behav-
ior behavior of the static potential, while the first term provides the standard
Coulomb piece at large distance. The distance scale being determined by the
new mass parameter $m$.\\
A simple guess that leads to the exchange of the behavior in the two different
regimes is the following
\begin{equation}
 \mathcal{L}=-\frac{1}{4}F_{\mu\nu}F^{\mu\nu}+\frac{1}{4}F_{\mu\nu}\frac{m^2}{\partial^2}F^{\mu\nu} \label{lw2}
\end{equation}

While the former Lagrangian (\ref{lw}) is an higher derivative local field theory,
plagued with all due complications 
\cite{Anselmi:2017yux,Anselmi:2018kgz,Donoghue:2018lmc,Sakoda:2019vhd}, 
the latter model (\ref{lw2})  is a non-local
theory. Although this term is added to the classical Maxwell Lagrangian ad
hoc, it represents a genuine quantum correction as will be shown later on.\\
Fortunately, the non-local term can be converted into a local one by the introduction of an additional field suitably 
coupled to $F_{\mu\nu}$.\\
In the extensions of $QED$ mentioned above one needs to introduce an arbitrary mass scale parameter $m$ to preserve 
the canonical dimension of the
gauge vector field $A_\mu$. In the wider framework of the Standard model of Elementary Particles masses are generated by 
the spontaneous breaking of local symmetries. Accordingly, we shall explore the possibility of generating $m$ in
the Lagrangian (\ref{lw2})  through a suitable Higgs mechanism.\\
This paper is organized as follows. In Section (\ref{lwed}) we discuss $QED$ coupled to
a massive Kalb-Ramond tensor field. We show that, in this model, the inter-
action potential between static charges is of Cornell type. We also define the
dual version of this model describing interacting open strings.
In Section(3) we generalize the model by coupling both fields to a pair of dis-
tinct scalar Higgs fields which generate, in alternating way, the mass for one
or another field. The resulting potential are either of Yukawa or Cornell type.
In Section (4) we summarize and discuss the obtained results.

\section{Kalb-Ramond $QED$}
\label{lwed}
We start from a quantum theory describing an Abelian vector field Aμ coupled
to a massive Kalb-Ramond anti-symmetric tensor $B_{\mu\nu}$ 
\cite{Smailagic:2001ch}. The euclidean
generating functional is given by

\begin{equation}
 Z \left[\, J \,\right] =Z \left[\, 0 \,\right]^{-1}
\int D \left[\, A\,\right] D \left[\, B\, \right] e^{-\int d^4 x\,\mathcal{L} }
\end{equation}
where the functional measure is defined to include the proper gauge fixing and
ghost terms; the Wick rotated Lagrangian density is:

\begin{eqnarray}
 &&\mathcal{L}= \frac{1}{4}F_{\mu\nu}F^{\mu\nu} -e J^\mu A_\mu +\frac{1}{12} H^{\mu\nu\rho}H_{\mu\nu\rho}
              +\frac{m^2}{4} B_{\mu\nu}B^{\mu\nu} -\frac{m}{2} F^\ast_{\mu\nu} B^{\mu\nu}\ ,\label{four}\\
 && F_{\mu\nu}=\partial_{[\,\mu} A_{\nu\,]}    \ ,\\
 && F_{\mu\nu}^\ast \equiv \frac{1}{4!}\epsilon_{\mu\nu\rho\sigma} F^{\rho\sigma}\ ,\\
 && H_{\mu\nu\rho}=\partial_{[\,\mu} B_{\nu\rho\,]}
\end{eqnarray}

$F_{\mu\nu}^\ast$  is the Hodge dual of $F_{\mu\nu}^\ast$ and $H_{\mu\nu\rho} $ is the Kalb-Ramond field strength.
The sign in front of the kinetic terms is determined in a way to provides the
correct damping factor in the functional integral.\\
On a general ground, the coupling constant in front of the $BF^\ast$ term need
not be the same as the mass m of the Kalb-Ramond field. However, it will
be shown that the desired Cornell potential can be obtained only if the two
parameters are identical.\\
$J^\mu$ is the divergence-free electromagnetic current density.
The action is at most quadratic in $A$ and $B$ by construction. Therefore, the
generating functional is Gaussian integral and can be evaluated by using classical field equations.\\
$\mathcal{L}$ is called parent Lagrangian in the language of duality \cite{Hjelmeland:1997eg}. 
Thus, integrating
out one or another field, one obtains two dual formulation of the same theory.\\
We are interested in calculating the classic, static potential between a pair of
test charges interacting via the exchange of Aμ quanta in the framework of an
effective $QED$.\\
To do so, we write down the classical field equations

\begin{eqnarray}
&& \partial_\mu \left(\, F^{\mu\nu} + \frac{m}{2}\epsilon^{\mu\nu\rho\sigma} B_{\rho\sigma}\,\right)= -e J^\mu\ ,\\
&& \partial_\lambda H^{\lambda\mu\nu} -m^2 B^{\mu\nu}= -m F^{\ast\,\mu\nu} \label{nove}
\end{eqnarray}

From equation (\ref{nove}) one obtains the divergence-free condition for $ B^{\mu\nu}$ 

\begin{equation}
 \partial_\mu B^{\mu\nu}= -\frac{1}{m } \partial_\mu  F^{\ast\,\mu\nu}\equiv 0
\end{equation}

 by virtue of the Bianchi Identities for $F_{\mu\nu}\left(\,A\,\right) $. Thus, equation (\ref{nove}) can be
written as

\begin{equation}
 \left(\, \partial^2 -m^2\,\right)B^{\mu\nu}  = - m  F^{\ast\,\mu\nu}
\end{equation}

and consequently

\begin{eqnarray}
 && B^{\mu\nu} = - \frac{m }{\partial^2 -m^2  } F^{\ast\,\mu\nu}\ ,\label{twelve}\\
 && H_{\lambda\mu\nu}= -m \partial_{[\,\lambda}\frac{1}{\partial^2 -m^2  } F^{\ast}_{\mu\nu\,]}\label{thirteen}
\end{eqnarray}

Inserting (\ref{thirteen}) and (\ref{twelve}) in $\mathcal{L}$, one finds an effective, non-local Lagrangian for $A_\mu$

\begin{eqnarray}
 \mathcal{L}_{eff}\left[\, A\,\right]&&= \frac{1}{4}F_{\mu\nu}F^{\mu\nu} -e J^\mu A_\mu 
+\frac{m^2}{4} F_{\mu\nu}\frac{1}{\partial^2 -m^2  } F^{\mu\nu}\ ,\nonumber\\
&&= F_{\mu\nu}\frac{\partial^2}{\partial^2 -m^2  } F^{\mu\nu}-e J^\mu A_\mu
\end{eqnarray}

The non-locality being a consequence of the $B F^\ast$ interaction in (\ref{four}). Integrating
out the gauge field in the generating functional, one finds

\begin{equation}
  \mathcal{L}_{eff}\left[\, J\,\right]=-\frac{e^2}{2} J^\mu \frac{\partial^2 -m^2  }{\left(\,\partial^2\,\right)^2}J_\mu
\end{equation}
We are now ready to obtain the classical potential energy between two test
charges. In detail, having two heavy static charges positioned in $\vec{x}_1$ and $\vec{x}_2$,
the only non-vanishing component of the current $J^\mu$ is

\begin{equation}
  J^\mu = e \delta^\mu_0 \left[\,\left(\, \vec{x}- \vec{x}_1\,\right)-\left(\, \vec{x}- \vec{x}_2\,\right) \,\right]
\end{equation}

The resulting \emph{effective action} reads

\begin{eqnarray}
S_{eff} &&=-\frac{e^2}{2} \int d^3 x \left\{\,\left[\,\delta^{3}\left(\, \vec{x}- \vec{x}_1\,\right)-
\delta^{3}\left(\, \vec{x}- \vec{x}_2\,
\right) \,\right]   \times  \right.\nonumber\\
 && \left.  \frac{-\nabla^2_x +m^2  }{\left(\,-\nabla_x^2\,\right)^2}
\left[\,\left(\, \delta^{3}\vec{x}- \vec{x}_1\,\right)-\delta^{3}\left(\, \vec{x}- \vec{x}_2\,\right) \,\right]\,\right\}
\label{17}
\end{eqnarray}

The cross terms in (\ref{17}) describes the potential energy $V_{int} \left(\, \vec{x}_1\ , \vec{x}_2\,\right)$ 
between the two charges as

\begin{equation}
 V_{int} \left(\, \vec{x}_1\ , \vec{x}_2\,\right)=-\int d^3 x\left[\,\delta^{3}\left(\, \vec{x}- \vec{x}_1\,\right)
\frac{-\nabla^2_x +m^2  }{\left(\,-\nabla_x^2\,\right)^2}\delta^{3}\left(\, \vec{x}- \vec{x}_2\,\right)\,\right]
 +\left(\,\vec{x}_1\leftrightarrow \vec{x}_2\,\right)
\end{equation}

We discard the divergent self-energy of each charge. Subsequent integration
over the dummy variable $\vec{x}$ leads to a Cornell potential

\begin{equation}
  V_{int} \left(\, r\,\right)= -\frac{e^2}{4\pi r} + \frac{e^2 m^2}{8\pi }r\ ,
\qquad r\equiv \vert \, \vec{x}_1\ - \vec{x}_2\,\vert  \label{19}
\end{equation}

Equation (\ref{19}) consists of two parts: the first describes the Coulomb interaction
dominant at short-distances, while the second part describes linear, confining
long distance tail reminiscent of the Kalb-Ramond interaction with the vector
potential. We stress that the $B F^\ast$ interaction term, with the mass $m$ as a
coupling strength, is instrumental for the proper modification of the standard
electrodynamics leading to the linear part of the potential.

\subsection{Dual ``~stringy~'' phase}

The usual interpretation of the confining phase has already been described in
terms of flux tubes connecting the particle/anti-particle pair. The dynamics
of such an object is usually described in terms of an open string with the two
charges at its end-points.\\
We shall show that such a description can be obtained by integrating out $A_\mu$
in favor of $B_{\mu\nu}$ in the parent Lagrangian $\mathcal{L}$. This gives the dual version of
the model described above.\\
The field equations for $A_\mu$ allow to express $F_{\mu\nu}$ in terms of $B_{\mu\nu}$

\begin{equation}
 \frac{\delta\mathcal{L} }{\delta A_\mu}=0\quad \Rightarrow \partial_\lambda \left(\, F^{\lambda\mu}- B^{\ast\,\lambda\mu}
\,\right) =- e J^\mu \label{20}
\end{equation}

The solution is obtained as follows. Firs, let us consider a current with support
on the world-line(s) of a real pair of opposite charges. Thus,the \emph{boundary}
current $J^\mu$ can be associated with a \emph{surface} current $\Sigma^{\mu\nu}$

\begin{equation}
 \partial_\mu J^\mu =0 \Longrightarrow J^\mu\left(\,x\,\right) =\partial_\lambda \Sigma^{\lambda\mu}\left(\, x\,\right) =
\partial_\lambda \int_\Sigma dz^\lambda\wedge dz^\mu \delta\left(\, x -z\,\right)
\end{equation}

where $\Sigma$ is the \emph{world-surface} swept by an open string connecting the two
opposite charges. With this relation between the two currents, equation (\ref{20})
has a solution

\begin{equation}
 F^{\mu\nu}= m B^{\ast\,\mu\nu} - e \Sigma^{\mu\nu} \label{22}
\end{equation}

Inserting (\ref{22}) into $\mathcal{L}$ gives the dual Lagrangian:

\begin{equation}
 \mathcal{L}_{KR}\left(\, B\ ;\Sigma\,\right)= \frac{1}{12} H_{\lambda\mu\nu}H^{\lambda\mu\nu} + \frac{em}{2} 
B^\ast_{\mu\nu} \Sigma^{\mu\nu} -\frac{e^2}{4}  \Sigma^{\mu\nu} \Sigma_{\mu\nu} \label{23}
\end{equation}

The Lagrangian (\ref{23}) describes a Kalb-Ramond tensor field $B_{\mu\nu}$ interacting
with the world-surface current $ \Sigma^{\mu\nu} $ of a string. The last term in  $ \mathcal{L}_{KR}$
is a Schild-type kinetic term for the string itself.\\
Integrating out the $B$ field in (\ref{23})  leads to the effective string Lagrangian

\begin{equation}
 \mathcal{L}_{eff}\left(\, \Sigma\,\right)= -\frac{e^2}{4}  \Sigma^{\mu\nu} \Sigma_{\mu\nu} 
 + \frac{e^2}{4}  \Sigma^{\mu\nu} \frac{m^2}{\partial^2}\Sigma_{\mu\nu} \equiv 
\mathcal{L}_{string}\left(\, \Sigma\,\right) \label{24}
\end{equation}

The dual picture translate the dynamics of single interacting charges via gauge
vector field into the interaction between open strings carrying opposite charges
at their end-points. This interaction is mediated by the exchange of a Kalb-
Ramond boson field.

\section{Kalb-Ramond Higgs mechanism}

As it has been shown in the previous Section, the non-local electrodynamics
encoded in the Lagrangian (\ref{lw2})  is equivalent to a local theory of interacting
gauge vector and massive Kalb-Ramond fields. Furthermore, it was necessary
to identify, a priori different, coupling constant of the interaction term with
the mass of the free Kalb-Ramond tensor in order to generate a confining
linear potential. It is generally argued that confinement is due to the different
vacuum state within hadrons with respect to the surronuding vacuum. This
suggests to use an Higgs mechanism in order to generate multiple vacua.
Also, the Higgs mechanism can naturally produce a single mass parameter in
the Lagrangian (\ref{lw2}). Thus, we introduce an additional interaction between the
Kalb-Ramond field and a neutral scalar field by the substitution

\begin{equation}
 mB_{\mu\nu}\longrightarrow g \psi B_{\mu\nu} \label{25}
\end{equation}

In order to give meaning to (\ref{25}) we assign a proper Higgs potential to $\psi$.
For the sake of generality, nothing forbids to introduce a complex scalar field
$\phi$ coupled to $A_\mu$. The resulting Lagrangian reads

\begin{eqnarray}
 \mathcal{L}&&= \frac{1}{4}F_{\mu\nu}F^{\mu\nu}  +\frac{1}{12} H^{\mu\nu\rho}H_{\mu\nu\rho} 
+\frac{g^2}{4}\psi^2 B_{\mu\nu}B^{\mu\nu} -\frac{g}{2}\psi F^\ast_{\mu\nu} B^{\mu\nu}-e J^\mu A_\mu\nonumber\\
&& +\frac{1}{2} \partial_\mu \psi \partial^\mu \psi +D_\mu^\ast \phi^\ast  D^\mu \phi -V\left(\, \phi^\ast \phi\ ,\psi^2\,\right)
\label{26}
\end{eqnarray}
Where the covariant derivative and the Higgs potential are
\begin{eqnarray}
 && D_\mu \phi \equiv \left(\, \partial_\mu -i e A_\mu \,\right) \phi\ ,\\
  && V\left(\, \phi^\ast \phi\ ,\psi^2\,\right)= \frac{\mu^2}{2}\psi^2 -\mu^2 \phi^\ast \phi +\frac{\lambda_\psi}{4!}\psi^4
+\frac{\lambda_\phi}{6}\left(\, \phi^\ast \phi\,\right)^2 
\end{eqnarray}
One can notice the alternation of sign in the quadratic terms. This choice
is made in order to have one or the other field developing a non-vanishing
vacuum expectation value, but not both at the same time. This will lead to
different phases corresponding to different vacua.
The charged Higgs field can be conveniently dealt with in the spherical basis
where the Lagrangian (\ref{26}) reads

\begin{eqnarray}
 \mathcal{L}&&=  \frac{1}{4}F_{\mu\nu}F^{\mu\nu}  +\frac{1}{2}\left[\,\partial_\mu\rho\partial^\mu \rho +
\rho^2 \partial_\mu\theta \partial^\mu\theta -2e^2 \rho^2 A_\mu \partial^\mu \theta +e^2\rho^2 A^2\,\right]\nonumber\\
&& +\frac{1}{2} \partial_\mu \psi \partial^\mu \psi -V\left(\, \rho\ ,\psi^2\,\right)+\frac{1}{12} H^{\mu\nu\rho}H_{\mu\nu\rho} 
\nonumber\\
&&+\frac{g^2}{4}\psi^2 B_{\mu\nu}B^{\mu\nu} -\frac{g}{2}\psi F^\ast_{\mu\nu} B^{\mu\nu}-e J^\mu A_\mu
\end{eqnarray}

\begin{equation}
 V\left(\,\rho^2\ ,\psi^2\,\right)= \frac{\mu^2}{2}\psi^2 -\frac{\mu^2}{2} \rho^2 +\frac{\lambda_\psi}{4!}\psi^4
+\frac{\lambda_\phi}{4!}\rho^4 
\end{equation}
The stationary points of the Higgs potential are given by

\begin{eqnarray}
 && \frac{\partial V}{\partial \rho}= -\mu^2 \rho +\frac{\lambda_\phi}{6}\rho^3\ ,\\
 &&\frac{\partial V}{\partial \psi}= -\mu^2 \rho +\frac{\lambda_\psi}{6}\psi^3
\end{eqnarray}

Solving the above equations one obtains three different vacua:

\begin{equation}
 A) : \mu^2=0 \qquad \rho_0=0\ ,\qquad \psi_0=0 \ ,\qquad \hbox{Coulomb vacuum}
\end{equation}

\begin{equation}
 B) : \mu^2>0 \qquad \rho_0^2=\frac{6\mu^2}{\lambda_\phi}\ ,\qquad \psi_0=0 \ ,\qquad \hbox{Yukawa vacuum}
\end{equation}

\begin{equation}
 C) : \mu^2<0 \qquad \rho_0=0\ ,\qquad \psi_0=\frac{6\mu^2}{\lambda_\psi} \ ,\qquad \hbox{Cornell vacuum}
\end{equation}

Since we are interested in the physical content of the theory in each phase
separately, we ignore quantum fluctuations and ”~\emph{freeze}~” the scalar fields in
their vacuum states. A similar phase portrait has been recently obtained 
a recent work  where, instead of a vector and a tensor, two different vector fields have been coupled
to a pair of Higgs fields \cite{Scott:2018xgo}.
\\
In each of the phases listed above the corresponding effective Lagrangian for the fields $A_\mu$ ,
$\theta$, and $B_{\mu\nu}$ is

\begin{eqnarray}
 \mathcal{L}&&=  \frac{1}{4}F_{\mu\nu}F^{\mu\nu}  +\frac{1}{2}\left[\,
\rho^2_0 \partial_\mu\theta \partial^\mu\theta -2e^2 \rho^2_0 A_\mu \partial^\mu \theta +e^2\rho^2_0 A^2\,\right]\nonumber\\
&& -V\left(\, \rho_0^2\ ,\psi^2_0\,\right)+\frac{1}{12} H^{\mu\nu\rho}H_{\mu\nu\rho} 
\nonumber\\
&&+\frac{g^2}{4}\psi^2_0 B_{\mu\nu}B^{\mu\nu} -\frac{g}{2}\psi_0 F^\ast_{\mu\nu} B^{\mu\nu}-e J^\mu A_\mu
\end{eqnarray}

As we do not include gravitational effects we can safely drop the constant term $V\left(\, \rho_0^2\ ,\psi^2_0\,\right)$.\\
Now, we can integrate out the Goldstone boson $\theta$ using its field equation

\begin{equation}
 \partial^2 \theta = e \partial_\mu A^\mu \longrightarrow \theta = e \frac{1}{\partial^2}\partial_\mu A^\mu
\end{equation}

leading to:

\begin{eqnarray}
 \mathcal{L}&&=  \frac{1}{4}F_{\mu\nu}\left[\, 1 - \frac{e^2\rho_0^2}{\partial^2}\,\right] F^{\mu\nu} -e J^\mu A_\mu \nonumber\\
\nonumber\\
&&+\frac{g^2}{4}\psi^2_0 B_{\mu\nu}B^{\mu\nu} -\frac{g}{2}\psi_0 F^\ast_{\mu\nu} B^{\mu\nu}+\frac{1}{12} H^{\mu\nu\rho}H_{\mu\nu\rho} 
\end{eqnarray}

The next step is to eliminate $B_{\mu\nu} $  using its equation of motion which expresses
it in terms of $F_{\mu\nu} $ as

\begin{equation}
 B_{\mu\nu}=- \frac{g\psi_0}{\partial^2 -g^2\psi_0^2} F_{\mu\nu}^\ast
\end{equation}

The final form of the gauge field Lagrangian is

\begin{equation}
 \mathcal{L}=  \frac{1}{4}F_{\mu\nu}
\left[\, 1 - \frac{e^2\rho_0^2}{\partial^2}+\frac{g^2\psi_0^2}{\partial^2 -g^2\psi_0^2} \,\right] F^{\mu\nu} -e J^\mu A_\mu 
\end{equation}

We shall now discuss the particular cases of interest.

\begin{itemize}
 \item Coulomb phase: $\mu^2 = 0 \rightarrow \rho_0 = 0\ , \psi_0 = 0$ this is the ”~trivial~“ vacuum
where both fields have vanishing expectation values and the theory reduces
to ordinary Maxwell electrodynamics

\begin{equation}
  \mathcal{L}=  \frac{1}{4}F_{\mu\nu}F^{\mu\nu} -e J^\mu A_\mu 
\end{equation}

\begin{equation}
  V_{int} \left(\, r\,\right)= -\frac{e^2}{4\pi r} 
\end{equation}

\item Yukawa phase: $ \mu^2 > 0 \rightarrow \rho_0 =6\mu^2/\lambda_\phi  \ , \psi_0 = 0$ In this case we obtain a
non-local Lagrangian for $A_\mu$

\begin{equation}
 \mathcal{L}=  \frac{1}{4}F_{\mu\nu}
\left[\, 1 - \frac{e^2\rho_0^2}{\partial^2} \,\right] F^{\mu\nu} -e J^\mu A_\mu 
\end{equation}

which leads to the non-local current interaction term

\begin{equation}
L\left[\, J\,\right]= - \frac{e^2}{2} \, J^\mu \frac{1}{\partial^2 -e^2\rho_0^2} J_\mu  \label{45b}
\end{equation}
For (\ref{45b}) one obtains the screened Coulomb potential, or Yukawa potential

\begin{equation}
  V_{int} \left(\, r\,\right)= -\frac{e^2}{4\pi r} e^{-e\rho_0 r}
\end{equation}

\item Cornell phase: $ \mu^2 < 0 \rightarrow \psi_0 =6\mu^2/\lambda_\psi  \ , \rho_0 = 0$  In this non-trivial vacuum
the vector gauge field and current Lagrangians are
\begin{equation}
 \mathcal{L}=  \frac{1}{4}F_{\mu\nu}
\left[\, 1 - \frac{\partial^2}{\partial^2-g^2\psi_0^2} \,\right] F^{\mu\nu} -e J^\mu A_\mu 
\end{equation}
\begin{equation}
L\left[\, J\,\right]= - \frac{e^2}{2} \, J^\mu \frac{\partial^2 -g^2\psi_0^2}{\left(\, \partial^2\,\right)^2} J_\mu  \label{45}
\end{equation}

leading to the linearly confining Cornell potential

\begin{equation}
  V_{int} \left(\, r\,\right)= -\frac{e^2}{4\pi r} + \frac{e^2 g^2\psi_0^2}{8\pi }r
\end{equation}
We mention that similar conclusions can also be obtained in a different ap-
proach based on the description of QED in terms of gauge invariant variables
\cite{Gaete:1998vr,Gaete:1999iy,Gaete:2007zn,Gaete:2007sj,Gaete:2009xf}.

\end{itemize}

\section{Summary and conclusions}

In this letter we have introduced a novel way to generate a confining linear
potential in the framework of an Abelian gauge theory. This model has been
chosen for its relative simplicity and also motivated by the shared belief that,
even in non-Abelian case, confinement is related to the Abelian sub-group.
The standard Maxwell theory, unavoidably, leads to a Coulomb potential. In
order to obtain a different behavior at a given range of distance one needs
to introduce a suitable modification. Podolski e Bopp, and later Lee-Wick,
introduced higher order derivatives terms in the Maxwell Lagrangian to ob-
tain a regular potential at short distance 
\cite{bopp,Podolsky:1942zz,Podolsky:1944zz}. Since we are looking for a
linear large distance potential, we introduced an inverse Lee-Wick type term
resulting in a non-local total Lagrangian. Non-localities are usually associated
with the quantum corrections. In the path-integral formalism these terms arise
after integration of on, or more, fields in the original Lagrangian.
In our case, the modified Lagrangian can be written in a local form by coupling
the gauge vector $A_\mu$ to a massive dynamical Kalb-Ramond tensor field $B_{\mu\nu}$ .
Written in the local form the physical content of the theory is transparent.
The interaction term has to be of the form mB F where the Kalb-Ramond
mass m is also the coupling constant.\\
This double role for m can be derived from a more fundamental interaction
of $B$ with a neutral Higgs scalar field $\psi$.
For the sake of full generality, we also coupled $A_\mu$ to a charged complex Higgs
field $\phi$. The choice of opposite signs in the quadratic terms in the Higgs potential allows to study 
two different phases in which only one of the two fields
develops a non-vanishing vacuum expectation value in a alternating way.
\\
In one phase the photon is massive and the Kalb-Ramond field remains massless and one obtains 
a screened Coulomb potential asymptotically approaching Yukawa form.\\
In the other phase, it is the photon which is massless while the Kalb-Ramond
becomes massive and the corresponding potential is of the Cornell type.\\
The emerging final picture describes confinement, though in a simplified set-
ting, as being induced by the interaction between the gauge and the Kalb-Ramond fields leaving in a non-trivial 
vacuum state generated through a Higgs mechanism.\\
We believe this model can offer useful hints of how to tackle confinement prob-
lem in a more realistic, but technically more involved, non-Abelian framework.

\end{document}